\documentclass[aps,apl,twocolumn,superscriptaddress,showkeys]{revtex4-1}

\usepackage{color}

\usepackage{graphicx}
\usepackage{dcolumn}
\usepackage{bm}
\usepackage{SIunits}
\usepackage[english]{babel}
\usepackage[latin1]{inputenc}
\usepackage[bottom=2cm,top=2cm, left=1.5cm, right=1.5cm]{geometry}

\usepackage[unicode=true, bookmarks=true, bookmarksnumbered=false, bookmarksopen=false, breaklinks=false, pdfborder={0 0 1}, backref=false, colorlinks=false]{hyperref}
\hypersetup{pdftitle={Infrared transmission spectroscopy of charge carriers in self-assembled InAs quantum dots under surface electric fields.}, pdfauthor={S. Pal, S. R. Valentin, N. Kukharchyk, H. Nong, A. B. Parsa, G. Eggeler, A. Ludwig, N. Jukam, A. D. Wieck}, pdfkeywords={Intersublevel spacings, Quantum dots, 2DEG, Epitaxial gates, FTIR transmission spectroscopy}}

\begin{document}

\title{Infrared transmission spectroscopy of charge carriers in self-assembled InAs quantum dots under surface electric fields}

\author{Shovon Pal}
\email{shovon.pal@ruhr-uni-bochum.de}
\affiliation{Lehrstuhl f\"{u}r Angewandte Festk\"{o}rperphysik, Ruhr-Universit\"{a}t Bochum, D-44780 Bochum, Germany.}
\affiliation{AG Terahertz Spektroskopie und Technologie, Ruhr-Universit\"{a}t Bochum, D-44780 Bochum, Germany.}

\author{Sascha R. Valentin}
\affiliation{Lehrstuhl f\"{u}r Angewandte Festk\"{o}rperphysik, Ruhr-Universit\"{a}t Bochum, D-44780 Bochum, Germany.}

\author{Nadezhda Kukharchyk}
\affiliation{Lehrstuhl f\"{u}r Angewandte Festk\"{o}rperphysik, Ruhr-Universit\"{a}t Bochum, D-44780 Bochum, Germany.}

\author{Hanond Nong}
\affiliation{AG Terahertz Spektroskopie und Technologie, Ruhr-Universit\"{a}t Bochum, D-44780 Bochum, Germany.}

\author{Alireza B. Parsa}
\affiliation{Institut f\"{u}r Werkstoffe, Ruhr-Universit\"{a}t Bochum, D-44780 Bochum, Germany.}

\author{Gunther Eggeler}
\affiliation{Institut f\"{u}r Werkstoffe, Ruhr-Universit\"{a}t Bochum, D-44780 Bochum, Germany.}

\author{Arne Ludwig}
\affiliation{Lehrstuhl f\"{u}r Angewandte Festk\"{o}rperphysik, Ruhr-Universit\"{a}t Bochum, D-44780 Bochum, Germany.}

\author{Nathan Jukam}
\affiliation{AG Terahertz Spektroskopie und Technologie, Ruhr-Universit\"{a}t Bochum, D-44780 Bochum, Germany.}

\author{Andreas D. Wieck}
\affiliation{Lehrstuhl f\"{u}r Angewandte Festk\"{o}rperphysik, Ruhr-Universit\"{a}t Bochum, D-44780 Bochum, Germany.}

\date{\today}

\begin{abstract}
We present a study on the intersublevel spacings of electrons and holes in a single layer of InAs self-assembled quantum dots (SAQDs) using Fourier transform infrared (FTIR) transmission spectroscopy without the application of an external magnetic field. Epitaxial, complementary-doped and semi-transparent electrostatic gates are grown within the ultra high vacuum conditions of molecular beam epitaxy to voltage-tune the device, while a two dimensional electron gas (2DEG) serves as back contact. Spacings of the hole sublevels are indirectly calculated using the photoluminescence spectroscopy along with FTIR spectroscopy. The observed spacings fit well to the calculated values for both electrons and holes. Additionally, the intersubband resonances of the 2DEG are enhanced due to the QD layer on top of the device.

\end{abstract}

\maketitle

 Quantum-confined zero-dimensional semiconductor quantum dots (QDs) have attracted an unparallel interest over more than a decade \cite{Faist2012,Ploog1990}. These nanostructures have an immense potential for device applications ranging from simple electronic memories \cite{Ritchie2009} to novel optoelectronics \cite{Bhattacharya2007}. The novelty can be extended by combining a quantum well (QW) with these zero-dimensional systems. Tunneling dynamics (refilling times) of electrons from QW to QD are tunable from $\micro\second$ to $\milli\second$ by changing the barrier thickness between them \cite{Faist2012,Faist2010}.  

 Read-out of memory devices based on SAQDs demands an effective electrical control over the device. One of the first experiments towards this goal was performed by Sakaki et al. \cite{Noge1995}, where they studied the transport properties of the 2DEG with embedded InAs QDs. A lot of work has been done to study the changes in conductance, carrier concentration and mobility of the 2DEG due to successive charging of QD levels \cite{Wieck2009,Geller2011}. Electrostatic quantization results in the formation of quasi-two-dimensional electron or hole states called subbands or sublevels. The spacing between the subbands is an important parameter defining device applications. Intersublevel spacings in InAs SAQD infrared photodetectors were calculated using infrared photoconductivity measurements \cite{Ploog1992,Petroff1994}. Besides, a lot of investigation has been carried out on the intersubband transitions of the 2DEG using grating couplers \cite{Schlapp1989}. However, questions arising on the influence of QDs on the intersubband spacings of the 2DEG and vice-versa in coupled nanostructures have not been addressed. Several theoretical models \cite {Bimberg1995,BimbergQD} are proposed to study the sublevel spacings. Nevertheless, due to lack of experimental agreement, determination of such energetic spacings still remains a subject of intense research.

 In this letter, we address these questions and study in depth the intersublevel spacings in QDs and its influence on the 2DEG intersubband spacing by FTIR transmission spectroscopy without applying an external magnetic field. This spectroscopy is done together with capacitance-voltage (CV) spectroscopy and photoluminescence (PL) spectroscopy. We also study the sublevel spacing of holes in the valence band of InAs SAQDs.
\begin{figure}[b!]
\includegraphics[width=\linewidth]{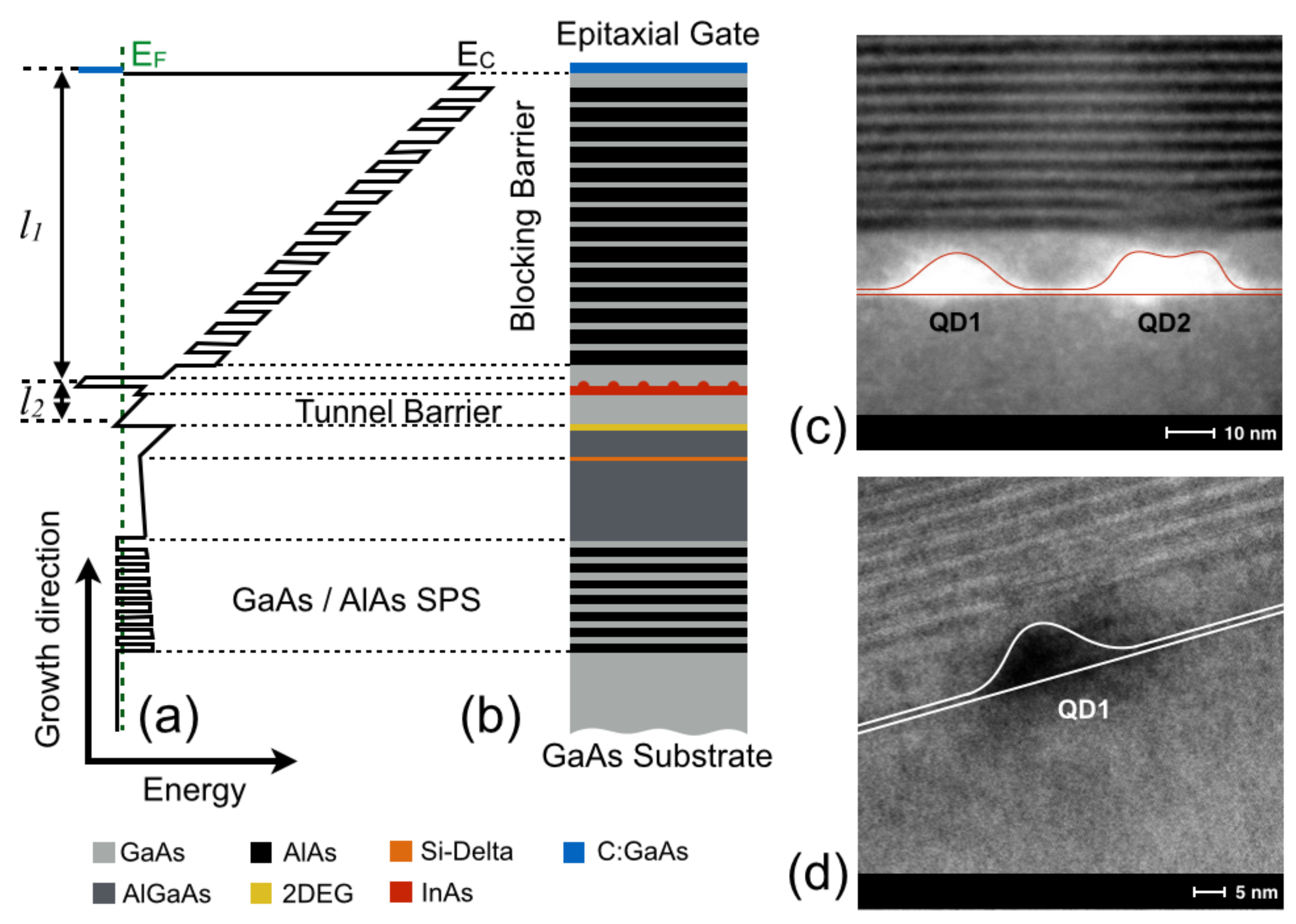}
\caption{\textbf{(a)} Sketch of the conduction band edge E$_c$, when no QD level is filled with electrons. $l_1$ is the distance of the QD from the gate (246 $\nano\meter$) and $l_2$ is the thickness of the tunnel barrier (30 $\nano\meter$). \textbf{(b)} Schematic of the layer sequence of the sample. \textbf{(c)} STEM image of the two QDs spaced apart by 44 $\nano\meter$. QD1 has a base diameter of 27 $\nano\meter$ and height of 8 $\nano\meter$ and QD2 has a base diameter of 30 $\nano\meter$ and height of approximately 8 $\nano\meter$. An imprint of strain relaxation is observed through the first three periods of SPS over QD2. \textbf{(d)} Dark field TEM image of QD1 with a resolution of 5 $\nano\meter$.}
\label{fig:device}
\end{figure}

 The investigated sample is grown on a semi-insulating GaAs (100) substrate using molecular beam epitaxy (MBE). Fig. \ref{fig:device}(b) schematically shows the layer sequence of the sample. An inverted high electron mobility transistor (iHEMT) structure is fabricated, on top of which InAs SAQDs are realized by Stranski-Krastanov growth mode with a nominal thickness of 2.2 monolayers (MLs), separated from the 2DEG by a tunnel barrier of a 30 $\nano\meter$ thick GaAs layer. The QDs are then capped by 8 $\nano\meter$ GaAs followed by a blocking barrier, which consists of 60 periods of AlAs/GaAs short-period-superlattice (SPS). An epitaxial, complementary-doped and semi-transparent electrostatic gate is grown on top of the sample within the ultra high vacuum conditions of the MBE. It is composed of a 25 $\nano\meter$ thick bulk carbon-doped GaAs layer, followed by 40 periods of carbon-delta-doped and carbon-doped GaAs layers.
\begin{figure}[b!]
\includegraphics[width=\linewidth]{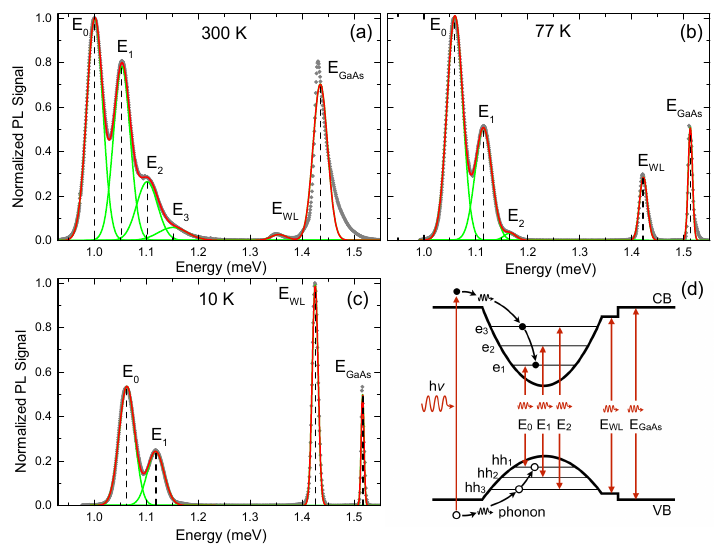}
\caption{\textbf{(a)} Room temperature, 300 $\kelvin$, \textbf{(b)} 77 $\kelvin$ and \textbf{(c)} 10 $\kelvin$ photoluminescence spectra measured with an excitation power of 5 $\milli\watt$. Spectral deconvolutions are plotted in green. The experimental data points are plotted in grey dots and the reconstructed spectrum is shown in red. \textbf{(d)} Schematic representation of the transitions corresponding to each peak.}
\label{fig:pl}
\end{figure}

 The carrier concentration of the epitaxial gate, obtained from Hall measurements, shows values of approximately $6 \times 10^{18}$ $\centi\meter\rpcubed$. There are a couple of advantages for using such gates. First, these gates have better optical transmission and do not suffer from low breakdown voltages in comparison to Schottky gates. Second, they grow lattice-matched and hence do not induce any strain on the underlying semiconductor layer, also verified from the scanning transmission electron microscope (STEM) studies. A sketch of the conduction band edge is shown in fig. \ref{fig:device}(a). The STEM image in fig. \ref{fig:device}(c) shows a cross-sectional view of typical QDs in the QD-ensemble of the device. Two QDs labeled as QD1 and QD2 show a variation in their shape. A visible strain is observed to relax through the first three SPS over QD2. The two dots are seperated by 44 $\nano\meter$, however the dot period in the device is random. An inverted etch-mask with an array of different dimensions ranging from $200 \times 200$ $\micro\meter$ to $500 \times 500$ $\micro\meter$ is used to prepare gates on top of the sample. The four corners of the sample are further etched by 200 $\nano\meter$ and Indium is diffused in an inert atmosphere of Hydrogen and Nitrogen to contact the 2DEG layer.
\begin{table}[b!]
\caption{Deconvolution results of the PL spectra at different temperatures. The values in the brackets are the FWHM of the respective peaks. All values are in $\electronvolt$.}
\centering
{\setlength{\tabcolsep}{8pt}
\begin{tabular}{l  c  c  c}
\hline\hline\\[-1.5ex]
Peaks & 300 $\kelvin$ & 77 $\kelvin$ & 10 $\kelvin$ \\ [0.5ex]
\hline\\[-1.5ex]
$E_0$  &  1.001 (0.036)  &  1.059 (0.036)  &  1.062 (0.036) \\
$E_1$  &  1.053 (0.038)  &  1.114 (0.038)  &  1.118 (0.038) \\
$E_2$  &  1.102 (0.047)  &  1.165 (0.027)  &  - \\
$E_3$  &  1.151 (0.066)  &  -  &  - \\
$E_{WL}$  &  1.355 (0.035)  &  1.422 (0.018)  &  1.425 (0.013) \\
$E_{GaAs}$  &  1.435 (0.033)  &  1.512 (0.009)  &  1.516 (0.006) \\ [1ex]
\hline\hline
\end{tabular}}
\label{table:deconvolution}
\end{table}
\begin{figure}[t!]
\includegraphics[width=0.9\linewidth]{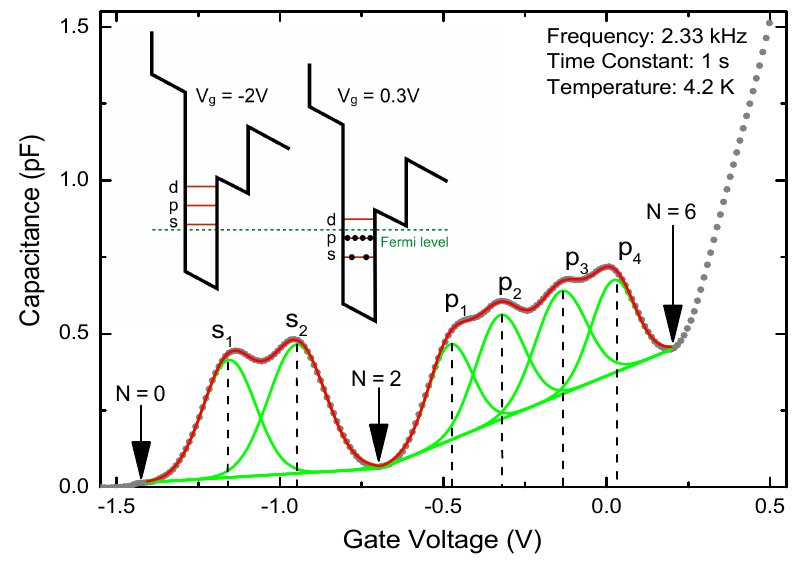}
\caption{Capacitance measured as a function of the gate voltage at 4.2 $\kelvin$. The capacitance is corrected by subtracting the linear slope due to the wetting layer. Successive charging of QD levels is observed in increasing the gate voltage. The green gaussian curves indicate the deconvoluted peaks corresponding to the filling of electrons in the QD levels. The red curve represents the reconstructed spectrum from $V_g$ $=$ -1.4 $\volt$ to $V_g$ $=$ 0.25 $\volt$. Inset: Band-schematic illustrates the QD levels with respect to the Fermi level for two gate voltages.}
\label{fig:cv}
\end{figure}
\begin{figure*}[t!h]
\includegraphics[width=\textwidth]{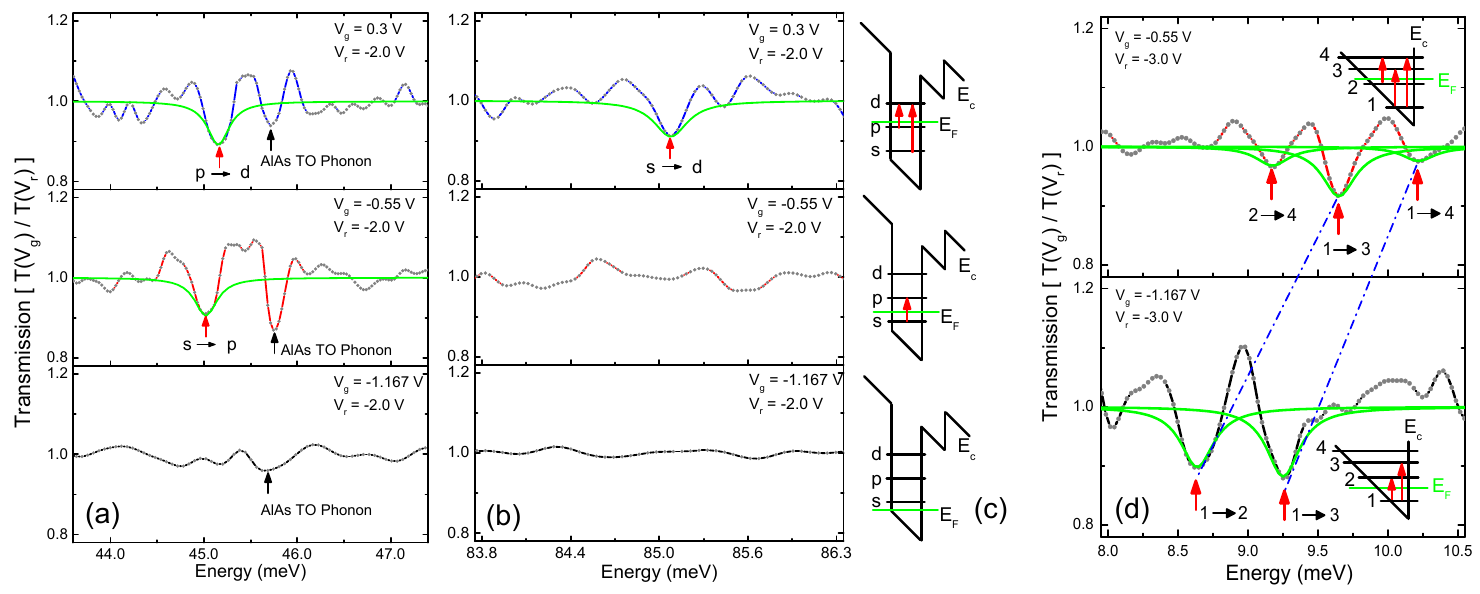}
\caption{Transmission spectra at three different gate voltages each normalized to the threshold voltage (no carriers induced) of -2 $\volt$.  \textbf{(a)} At -1.167 $\volt$, only an absorption due to AlAs TO-phonon (45.68 $\milli\electronvolt$) is observed. At -0.55 $\volt$, $s \to p$ transition is recorded at 45 $\milli\electronvolt$ along with the enhanced TO-phonon absorption at 45.74 $\milli\electronvolt$. At 0.3 $\volt$, $p \to d$ transition was observed at 45.14 $\milli\electronvolt$ along with the TO-phonon absorption at 45.71 $\milli\electronvolt$.  \textbf{(b)} Transmission spectra at higher frequencies than in (a): At 0.3 $\volt$, $s \to d$ transition was recorded at 85.07 $\milli\electronvolt$. \textbf{(c)} The level schemes for the respective transitions observed in (a) and (b). \textbf{(d)} Optically enhanced intersubband transitions in the 2DEG for two gate voltages at substantially lower frequencies than that of QDs in (a) and (b). The resonances shift to higher energies as the carrier density increases with the gate voltage. This is shown by the blue dashed lines. The green curves in all the spectra represent the simulated transmission spectra.
}
\label{fig:ftir}
\end{figure*}
Intraband spacings of the sublevels of electrons and holes in the InAs SAQDs (shown in fig.~\ref{fig:pl}) are characterized by photoluminescence measurements at three different temperatures. A modulated laser diode, which emits at 638 $\nano\meter$ (1.94 $\electronvolt$), is used to excite the sample and a liquid-nitrogen-cooled InGaAs photodiode is used as a detector. The photo-generated electrons and holes undergo energy and momentum relaxation towards the band-gap minima, where they finally recombine and emit photons with energies given by \cite{BimbergQD}
\begin{equation}\label{eq:eqn1}
{E_i} = {E_g} - {E_{{e_{i + 1}}}} - {E_{h{h_{i + 1}}}},
\end{equation}
where $i$ represents the index of the levels, $E_g$ is the band-gap of the matrix material (GaAs), $E_{{e_{i + 1}}}$ and $E_{h{h_{i + 1}}}$ are the energetic distances in the conduction and valence band measured from the respective edges of the GaAs band. Fig.~\ref{fig:pl}(d) shows the schematic of the intraband transitions in the QD. At 300 $\kelvin$, four intraband transitions are observed (fig.~\ref{fig:pl}(a)), marked as $E_0$ (1.001 $\electronvolt$), $E_1$ (1.053 $\electronvolt$), $E_2$ (1.102 $\electronvolt$) and $E_3$ (1.151 $\electronvolt$) along with the transition in the wetting layer (1.355 $\electronvolt$) and in GaAs (1.435 $\electronvolt$). The composition of this In$_{0.05}$Ga$_{0.95}$As wetting layer is calculated from the observed peak in PL (at 300 $\kelvin$) using Vegard's law \cite{Barns1978}. With decreasing the temperature to 77 $\kelvin$, the respective transitions shifted to higher energies (fig.~\ref{fig:pl}(b)) according to Varshni's emperical relation \cite{Varshni1967}. Further decreasing the temperature to 10 $\kelvin$, only two distinct intraband transitions are visible (fig.~\ref{fig:pl}(c)), marked as $E_0$ (1.062 $\electronvolt$) and $E_1$ (1.118 $\electronvolt$), together with the transition due to the wetting layer (1.425 $\electronvolt$) and GaAs (1.516 $\electronvolt$). The values of all the deconvoluted peaks at different temperatures along with their FWHM are tabulated in tab.~\ref{table:deconvolution}. The FWHM of $E_0$ and $E_1$ are found to remain constant at all temperatures while a marked narrowing of FWHMs of other peaks is observed at 10 $\kelvin$. The broadening of the peaks is due to the ensemble of QDs resulting in a series of $\delta$-function peaks whose position depend on the energy levels in the 3D-confined region. There is a strong emission from the wetting layer at 10 $\kelvin$ as compared to the emission at 300 $\kelvin$, which is also observed as a huge change in capacitance in the CV spectrum for gate voltages higher than 0.5 $\volt$. Two intraband transitions, $E_2$ and $E_3$, are not strong enough and hence not observed at 10 $\kelvin$.

 On varying the gate voltage ($V_g$) from -2 $\volt$ to 1 $\volt$, we inject electrons from the back-contact (2DEG) into the InAs QDs through the GaAs tunnel barrier. The charging of the QD levels with an electron is observed as a change in the capacitance between the top gate and the back contact. Fig.~\ref{fig:cv} shows the variation of capacitance with the gate voltage at 4.2 $\kelvin$. A small sine signal of 10 $\milli\volt$ and 2.33 $\kilo\hertz$ was used as a modulation for the DC voltage. In the inset of fig.~\ref{fig:cv}, schematics of the QD levels are shown for two different voltages. At -2 $\volt$, no QD levels are occupied and hence this voltage is chosen as reference voltage ($V_r$). With sufficient forward bias (0.3 $\volt$), we see the occupation of s-like $E_0$ level, which is two-fold degenerate and p-like $E_1$ level, which is four-fold degenerate. At an intermediate voltage of -0.55 $\volt$, only the s-like $E_0$ level is occupied. The density of QDs can be obtained by using a simple relation:
\begin{equation}
{N_{QD}} = \frac{1}{{2e}} \cdot \frac{{{l_1} + {l_2}}}{{{l_2}}}\int\limits_{ - 1.5}^{ - 0.7} {C\left( V \right)dV},
\end{equation}
where $N_{QD}$ is the density of QDs, $({{l_1} + {l_2}})/{l_2}$ is given by the lever-arm rule (see fig.~\ref{fig:device}(b)). The integration is performed between the voltage limits where only the s-levels are occupied. The density of the QDs calculated from the CV spectrum is $2 \times 10^9$ $\centi\meter\rpsquared$.

 Further, we investigate the quantized energy levels within the conduction band by infrared transmission measurements with a rapid scan Bruker IFS113V interferometer, which has two different sources (mercury-arc lamp and globar) to cover the complete range of infrared radiation. The spectral resolution was 0.25 $\reciprocal{\centi\meter}$ ($=$ 0.03 $\milli\electronvolt$). The bottom surface of the sample is wedged at around 3$\degree$ to avoid the Fabry-Perot interference fringes and mounted at an angle of 30$\degree$ (obeying the polarization selection rule for observing the intersubband transitions) in a self-built cryogenic optical sample holder, which is equipped with a liquid-Helium cooled Si-bolometer for detection. Spectra at a certain gate voltage and reference voltage are collected alternatively and averaged over long measurement times (3 hours) to rule out the experimental drifts. The normalized transmission, $T\left( {{V_g}} \right) /T\left( {{V_r}} \right)$, is given by \cite{Tsui1981}:
\begin{equation}\label{eq:transmission}
\frac{{T\left( {{V_g}} \right)}}{{T\left( {{V_r}} \right)}} = 1 - \frac{{2{\rm{Re}}\left[ {\sigma \left( \omega  \right)} \right]}}{{\left( {1 + \sqrt \varepsilon   + {r_v}/{r_g}} \right){\varepsilon _0}c}},
\end{equation}
where $\varepsilon_0$ is the permittivity of free space, $c$ is the velocity of light, $\varepsilon$ = 12 is the dielectric constant of InAs, $r_v$ = 377 $\ohm$ is the vacuum impedance and $r_g$ = 12 $\kilo\ohm$ is the combined impedance of the epitaxial gate and the back-contact at 4.2 $\kelvin$. The high-frequency conductivity, $\sigma \left( \omega  \right)$, of electrons in a parabolic potential is
\begin{equation}\label{eq:conductivity}
\sigma \left( \omega  \right) = \frac{{N{e^2}\tau }}{{2{m^*}{a^2}\left[ {1 + i\frac{{\left( {{\omega ^2} - \omega _r^2} \right)}}{\omega }\tau } \right]}},
\end{equation}
where $\omega _r$ is the high-frequency resonance, $a$ is the average QD periodicity and $\tau$ is the scattering time, which basically determines the line-width of the transitions. Theoretical curves generated by using eq.~\ref{eq:transmission} show close correspondence to the experimental data. These are shown in fig.~\ref{fig:ftir}. At $V_g$ = -1.167 $\volt$, we observe only a small dip in the transmission due to the AlAs TO-phonon at 45.7 $\milli\electronvolt$ \cite{Baroni1991}. At $V_g$ = -0.55 $\volt$, electrons from the s-like $E_0$ level absorb the incident infrared radiation and occupy the next sublevel with a minimum in the transmission spectra at 45 $\milli\electronvolt$ ($s \to p$). Upon increasing the voltage to 0.3 $\volt$, when both s- and p-levels are occupied, transitions from both the levels to the d-sublevel are registered as minima in the transmission spectrum at 45.1 $\milli\electronvolt$ ($p \to d$) and 85.1 $\milli\electronvolt$ ($s \to d$), respectively. For $s \to d$, the value is almost twice the value observed in $s \to p$ or $p \to d$. This is expected to a very good approximation for confinement resulting from the parabolic potential leading to equidistant levels. We observe a strong coupling of $s \to p$ and $p \to d$ to the AlAs TO-phonon (45.7 $\milli\electronvolt$). Using the results from the PL spectra, we calculate the sublevel spacings of the holes in the valence band. The energetic distance between hh$_1$ and hh$_2$ (see fig.~\ref{fig:pl}(d)) is 11 $\milli\electronvolt$.

 On the other hand, strong intersubband resonances (ISRs) are observed in the 2DEG. The 2D electron density obtained from Hall measurements is $3 \times 10^{11}$ $\centi\meter\rpsquared$. Sufficient negative bias ($V_r$ = -3 $\volt$) is applied to completely deplete the 2DEG below the threshold voltage, which is used as reference for the chopping scheme. The strong enhancement of the 2DEG-ISRs is due to the layer of QDs on top, which helps in better coupling of the incident radiation with the intersubband transitions of the 2DEG. This is due to the fact that the InAs SAQDs in the GaAs matrix act analogous to a grating coupler, which tilts locally the electric field vector of the incident wave, leading to substantial components in the growth direction (normal to the 2DEG), which is necessary to excite ISR. As a consequence, we observed around 6$\%$ transmission dip in the primary ISR compared to the strengths of the transitions reported previously 1-2$\%$ \cite{Schlapp1989}. With an increase in the gate voltage, the ISR shifts to higher energies owing to an increase in the carrier density steepening up the asymmetric triangular potential, which confines the 2DEG and thus leading to higher intersubband spacings.

 The picture is however not so simple since many-electron interactions significantly change the single electron intersubband transition energies of the 2DEG in a HEMT structure. Under the local density approximation and linear response theory, it was shown by Ando \cite{Ando1977} that the two most significant contributions are the depolarization effect which blue-shifts the ISR and the exciton-like effect which red-shifts the ISR. Besides, with an increase in the voltage (-0.55 $\volt$), the QDs are also charged and this will result in an additional field across the 2DEG layer. The resonances observed are a result of the combined effect of the above mentioned phenomena. Moreover, as seen from the cross-sectional STEM images, the QDs have an inhomogeneous size distribution and there are also signatures of strain relaxation through the SPS. It is quite remarkable to note that under the framework of parabolic potential (in case of QDs), the sublevel spacings are independent of the electron density. This is however true only for lower densities \cite{Gossard2001}, which is the case for the sample investigated.

 In conclusion, the use of epitaxial, complementary doped and semi-transparent electrostatic gates to chop the density of charge carriers proved to be an efficient technique to study the intersublevel spacings of electrons in the conduction band. Using these results along with the intraband transitions values from PL, we evaluated the sublevel spacings of the holes in the valence band. We found that the simulated transmission spectra correspond to a very good approximation to the experimental observations. Due to the coupled structure with the QD layer on top, the 2DEG intersubband resonances were strongly enhanced. Moreover, there was no visible coupling of the intersubband resonances of the 2DEG and the intersublevel transitions in the QDs and they are well seperated by the Reststrahlen band of GaAs. Interesting features of SAQDs like inhomogeneous confining potential resulting from the shape and strain relaxation were seen in the STEM images.

This work was financially supported by IMPRS-SurMat, MPIE D\"{u}sseldorf, BMBF Quantum communication program, BMBF-QUIMP 16BQ1062. The authors would also like to acknowledge the DFH/UFA CDFA-05-06 Nice-Bochum and the RUB Research School.

\end{document}